\documentclass[article,twocolumn]{revtex4}

\usepackage{amssymb,amsmath,amsfonts}
\usepackage{graphicx}

\begin{document}

\title{Understanding Vacuum Arcs and Gradient Limits}

\author{J. Norem$^*$}
\affiliation{Argonne National Laboratory, Lemont IL, USA,(retired)}
 \email{norem.jim@gmail.com}
 
\author{Z.  Insepov}
\affiliation{Nazarbayev University, Nur-Sultan, Kazakhstan}
\affiliation {National Research Nuclear University (MEPhI), Moscow, Russia}
\affiliation{Purdue University, West Lafayette, IN, USA}

 \author{A. Hassanein}
\affiliation{Purdue University, West Lafayette IN, USA}
 
\date{\today}
 
\begin{abstract}
Although a general model of vacuum arcs and gradient limits would be widely useful, roughly 120 years after the first good experimental data on these arcs, this important field continues to be unsettled.  This problem is a limitation in a number of technologies and has applications in many fields.  Large tokamaks are sensitive to arcing on the plasma facing components, linac costs depend on their maximum operating fields, power transmission efficiency depends on the voltage that can be maintained, and the efficiency of Atom Probe Tomography depends on avoiding sample failures.  A multidisciplinary study of this field could improve the precision and applicability of the theoretical models used.   We outline the basic mechanisms involved in arcing and the issues that determine the physics of arcs.  In order to look at the physical principles involved, we divide the process into four stages; the trigger, plasma formation, plasma evolution and surface damage.   We try to identify the dominant mechanisms, critical issues and desirable aspects of an R\&D program to produce a more precise and general model.

\end{abstract}

\maketitle

\section{Introduction}                       

The basic physics that determines the maximum accelerating fields in accelerators is not settled science.  Although vacuum arcs have been under active study for over 120 years, since the first precise experiments done by A. A. Michelson and R. Millikan at the University of Chicago in 1900$-$1905, the mechanisms involved in vacuum breakdown, arc physics and surface damage have been inconclusively debated \cite{earhart, Hobbs, Juttner}.  In part this is due to the fact that vacuum arcs are in constant use, and the primary goal in most applications is to produce working devices that are efficient, safe, reliable and inexpensive, and most R\&D has been directed at these goals.  In addition, arc physics is complex, due to the wide parameter ranges, high speeds, unpredictability, small dimensions, large number of applicable processes, non-linear and discontinuous mechanisms involved \cite{C, A, H, alpert, dyke, ro, dj, feynman}.  For example, we find that breakdown in RF systems may be dependent on duty cycle, space charge and pulse length \cite{nature}.

The goal of arc R\&D should be a single, numerical model that explains all stages of the behavior of vacuum arcs in a self-consistent way, from breakdown, to arc formation, arc evolution and surface damage and asperity creation \cite{IN}.  This is complicated somewhat by the wide variety of environments these arcs exist in, but simplified by other constraints, for example the limited range of materials are considered for both normal and superconducting systems.  

While the ultimate gradient limits of these materials are fairly well understood experimentally, we believe that the importance of vacuum arcs as a fundamental technological limit in many fields justifies continued multidisciplinary  study.   

\section{History}                   

Breakdown of gasses between two electrodes was studied by many in the period 1850$-$1897, eventually leading to the model of Townsend, based on an electron ionization avalanche \cite{jst}.  The logical next step, what would happen if there was no gas to break down between the electrodes, was then widely discussed.  Although the university had no vacuum pumps, Michelson, with the interferometers he had developed, was able to look at electrical breakdown over distances that were small compared to the ionization length and thus eliminate the contribution from the gas, and study breakdown of metallic surfaces at high fields \cite{earhart, Hobbs}.  Lord Kelvin, doing atomic physics without atomic models, produced a model of breakdown in 1905, based on tensile strength of materials, tensile stress produced by electric fields, and geometrical enhancements of electric field \cite{lordk}.

While the physics of vacuum arcs has been studied continuously since 1900, the wide range of initial conditions tended to produce a wide range of disagreement about the dominant mechanisms involved .  For example, pins in positive and negative electric fields broke down in completely different ways, depending somewhat on pulse lengths and geometry, thus different models have been used in different applications.  Important ideas that developed over the years, include field enhancements, Explosive Electron Emission (EEE) \cite{A}, the dependence on local geometry \cite{feynman}, and the various ways atomic structures respond to stresses.  In the case of RF  accelerators, explaining experimental data requires explaining how the breakdown occurs quickly, with sharp thresholds in clusters of breakdown events.

\section{Modeling}                   

We have found it useful to divide the process of arcing into individual stages, and then identify the dominant mechanisms in each stage of the evolution of the arc (which can depend on a variety of conditions), working towards a model in which all the stages can be modeled  in a self-consistent way \cite{IN}.  In general, this is fairly straightforward considering plasma properties, but more difficult when considering surface damage and subsequent breakdown events.  We have identified the following contributing mechanisms and stages:

{\bf Surface failure}:  Surface failure can occur either due to tensile stresses or exploding wire heating.

{\bf Ionization}:  In RF systems this process seems to require on the order of 10 ns to produce dense plasmas from the initial ionization of small volumes of material, by field emitted currents.

{\bf Plasma evolution}:  The plasma density rises to fixed levels, melting the surface and producing instabilities which limit the density.

{\bf Surface damage}:  Capillary waves smooth the surface and differential cooling can produce stresses that can roughen it.

\begin{figure}[htb]  
   \centering
   \includegraphics[width=7cm]{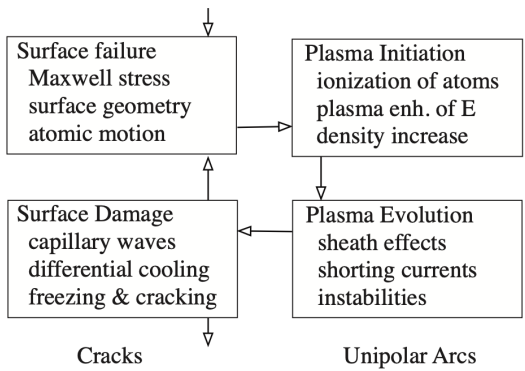}
   \caption{Vacuum arc development involves 4 stages \cite{IN}.  We consider processes that seem dominant at different stages of the development of the arc, and find that under continued operation the arc follows a life-cycle, where damage from one breakdown event is very likely to produce another.}
\end{figure}

This general model seems to be inclusive enough to be applicable to all accelerator applications and could be a useful guide or starting point for other applications.  We find that all the mechanisms used in our modeling become active at surface fields of around 10 GV/m, and it is sometimes difficult to determine which is dominant.

Historically, breakdown modeling was done for long pulse and DC applications with needle shaped samples providing local field enhancements.  When the surfaces were negatively charged, field emitted currents Ohmicly heated the samples until they exploded, ultimately producing a large literature on exploding wires and Explosive Electron Emission (EEE) \cite{A}.  On the other hand, if the surfaces were positively charged, the surface could explode with a Coulomb explosion, or slowly erode due to field evaporation, at a somewhat larger field.  Near breakdown threshold, the exploding wire model was significantly slower because of the time required to heat material to, and beyond, its melting point \cite{A}.

In RF applications, we find a number of complications to this picture.  Since field emission only occurs when the surface is charged negatively and the field emission current is, $I_{FE}  \sim E^{13}$, we find that heating is proportional to $I_{FE}^2 \sim E^{26}$ \cite{FN,PR1}.  Thus, the duty cycle for heating is reduced by a factor of roughly 13 from the DC case.   There is little experimental information on the geometry of the asperities that are breaking down, however they seem to be very small, and the smaller they are, the faster they would cool.  In one example, a right-angle corner would cool in $\sim 10^{-14} $ sec.

In addition to duty cycle problems, RF systems seem to be vulnerable to the space charge limit which would also limit the surface field and the maximum current produced at sharp corners, further reducing the heat produced \cite{dyke, IN}. 

We assume that the initiation of the arc is due to field emitted currents ionizing solid or gaseous material from the surface over a number of RF cycles.  In our simulations we find that this process takes around 10 ns to produce the level of current that can be detected externally \cite{Noremrf2011, dj}.

We assume that a unipolar arc is produced, whose parameters are determined by the applied external fields and the interactions of the plasma with the surface\cite{Schwirzke91}.  We assume the surface is quickly melted and electric fields disturb the surface, producing a turbulence and some particulate ejection.  We find that the surface field beneath the arc is also around 10 GV/m, which continues the field emission process \cite{morozov}.

When the external fields die off, the plasma will cool and terminate.  The surface will also begin to cool.  We find that the strong surface tension of the liquid metal will cause capillary waves and smoothing of the liquid surface, which produces the smooth surface which characterizes arc damage in metals \cite{LL}.  We have shown that the surface solidifies from the outside of the arc boundary and the surface contracts as it cools, which can produce cracks near the arc center \cite{IN}.   The widths of these cracks are comparable to that expected from thermal contraction.  We find that the many cracks and crack junctions produce large numbers of right-angle corners which would produce field emission and breakdown if high fields were subsequently applied.  Corners could be sharp on an atomic scale, producing high field enhancements without being particularly conspicuous and microroughness will also complicate estimates of field emission currents \cite{feynman}.

\begin{figure}[htb]  
   \centering
   \includegraphics[width=8cm]{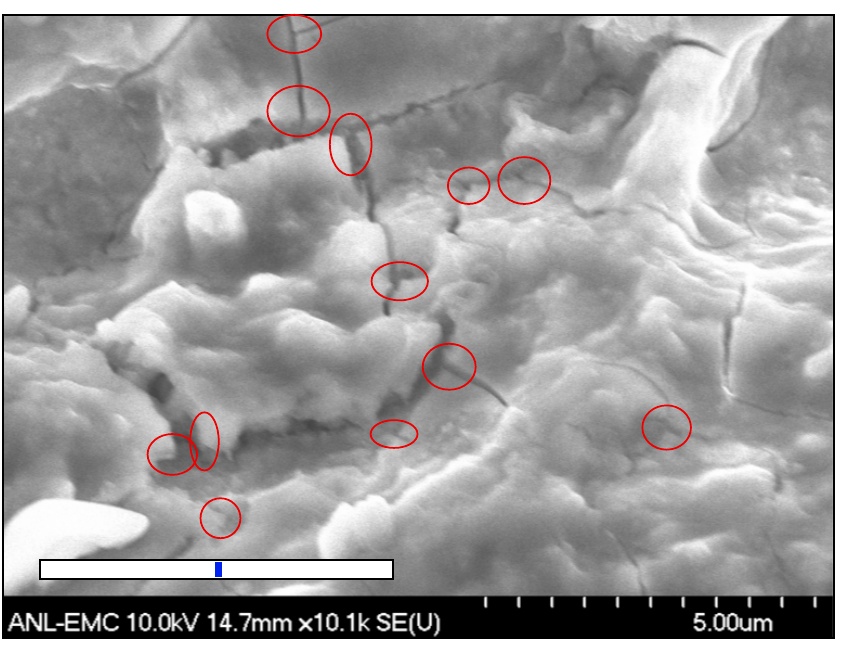}
   \caption{Many cracks visible in SEM images of the center of an arc damage area at a magnification of 10,100X.  Crack junctions where high field enhancements are expected, are noted. The widths of the cracks are caused by the thermal contraction of the material, $\Delta x \sim x \alpha \Delta T \sim$ 2\% of the initial section, as it cools after solidifying, where $\alpha$ is the coefficient of thermal expansion and $\Delta T$ is the change in temperature. The overall diameter of the damage is $\sim$ 500 $\mu$m, which explains the wider cracks.  The blue spot is $\sim$2\% of the length the white line for comparison. }
\end{figure}

Many effects occur around 10 GeV/m and determining which mechanism is dominating at any given time is primarily a question of which produces the most self-consistent overall model.  The following mechanisms seem to be important.

Field Emission:  Years of data show a dependence of the current, I, on the applied electric field proportional to $E^{13}$, consistent with detailed predictions of the Fowler-Nordheim model \cite{FN, H, PR1} when a total emitter area, duty cycle, cavity geometry and local field enhancements are considered.  Our data was consistent with theory over 14 orders of magnitude. 

Exploding wire physics and Explosive Electron Emission have been studied for years, and are the dominant mechanism in DC cathode arcs \cite{A}.  Short pulse, RF systems, however, seem to break down faster than the required heating times \cite{nature}.           

APT Surface failure and evaporation:  Atom Probe Tomography (APT) uses high surface fields to cause atoms to evaporate off polished surfaces \cite{F}.  When it works, it produces incredible three-dimensional images of the atomic structure of materials.  Many samples fail due to electrostatic discharges however, and when they fail all data is flushed.  APT references do not consider sample failures either theoretically or experimentally, however it seems to occur at surface fields of $\sim$30 GV/m for Cu for cold samples (20 - 80K).  Strength of materials is reduced at higher temperatures, and RF systems may involve fatigue and other effects which could lower the discharge threshold.

Electromigration: Electromigration is one of the most important constraints in the design of integrated circuits, however it is well understood \cite{D, DD}.  (Billions of iphones, each with billions of transistors, do {\bf not} fail because of this mechanism.)   We assume that this mechanism dominates diffusion during breakdown.  Field emitting surfaces seem to operate at current densities near the limits used in microelectronics, which would cause the surface microgeometry to evolve at rates $\propto i_{FE}^2 \sim E^{26}$, consistent with experiment \cite{ro}.  
	
Space charge: Analysis done in the 1950s has shown that space charge will limit surface fields and currents produced by electron emitters \cite{dyke}.  The local fields involved, $\sim$10 GV/m, fall into the narrow range where surface failure occurs.  We have modeled space charge effects during field emission at 90$^o$ corners, showing clouds of electrons drifting towards the surface due to space charge \cite{IN}.
	
Surface fields/enhancement factors: Most calculations and experimental work assume that the surfaces producing field enhancements involve some sort of needle/fencepost geometry, although they are not seen.  Feynman, in a short note, points out that the size of asperities is more important than their shape \cite{feynman}.
	
B fields: Collinear B fields can be used to alter the breakdown conditions in known ways, produce beamlets that produce images of field emission, and simplify the geometries involved with surface damage \cite{IN}.

\section{Critical Issues}                   

While these mechanisms have all been studied independently, the most interesting step is to combine the known mechanisms and numerically model the development of the arc throughout all the stages of development of the arc and surface structure with consistent surface geometry showing, for example, the production and breakdown of asperities, heating and cooling of the surface, ionization and termination of the unipolar arc and dependence of these mechanisms on the external parameters of the system.  There are a number of issues that must be resolved in order to simplify models of breakdown, extend them to other applications and parameter ranges and improve their predictive power, in addition to effects like duty cycle.  In general, plasma properties in cathodic arcs have been well studied due to its usefulness in sputter coating \cite{C}. 

{\bf Production of surfaces with significant enhancement factors} can be qualitatively explained assuming thin, metallic pools cool quickly from the outside, solidify and then continue to cool and shrink, causing surface cracks and crack junctions \cite{LL, IN}.  The cooling of the surface after the arc is terminated has not been studied in detail, although it seems to be a primary mechanism in surface damage.  Other mechanisms are also possible, including particulate production, but have not been modeled.  It is useful to consider applications with collinear B fields which seems to simplify the geometries without changing the physics.   

There is comparatively little {\bf experimental data on the geometry of asperities} capable of producing breakdown.  We can show corners experimentally produced at crack junctions, however there is little evidence of wire shaped asperities in the literature.   The shape and dimensions of asperities would determine the rates of breakdown and the required times for heating.

{\bf Space charge limited effects} have not been numerically modeled in a self-consistent way for realistic asperities.  This subject was actively studied in the 1950s by Dyke et. al. \cite{dyke, IN}, and this work may provide the most relevant approach in the literature. The subject is complicated by the $E^{13}$ dependence of current on surface field, the unknown local geometry of the emitters and other issues.  The space charge limit could explain the production of hollow beamlets we have seen, if emission from the tips of emitters was reduced, attenuating the intensity at the center of the beamlet \cite{IN}.
	
Even the basic dimensions and parameters of the arcs are poorly understood.  Surface instabilities beneath the plasma are possible, along with particulate production and acceleration.  While many mechanisms seem to be able to increase the plasma density, the mechanism that limits the density is not clearly identified \cite{C}.  Likewise the production and acceleration of macropartiles \cite{C}, the production of hollow beams \cite{PR1}, the field dependence of corona currents on power lines \cite{EPRI} and other effects are poorly explained.

One simplification seems to be the limited number of materials used in RF systems.  Most applications use copper, a well understood material, although the use of Be has been proposed in muon cooling systems.  Superconducting systems, which use a variety of materials that are subject to breakdown.

\section{Applications}                   

The interactions of plasmas and high fields with surfaces are active, important and not settled.  Although our primary interests are with RF acceleration of particles, this work is relevant to many problems, and related to many failure modes of electronic materials and devices \cite{D}. 

 In tokamak research, arcing can be very important to operational performance and is actively studied for dust, erosion,  plasma contamination and plasma transients \cite{Tanabe}.  In accelerator applications, the community has learned to operate systems with high gradients and good reliability without a complete understanding the physics, however extrapolation to higher RF frequencies, plasma and laser driven systems are still untested.  Corona losses and gradient limits on the power grid directly affect the operating voltage of these systems and partially determine the economic cost of power transmission \cite{corona, EPRI}.  Although current density limits affect the design of microelectronics, they are well understood and this experience should be incorporated into accelerator modeling \cite{DD}.  Micrometeorite impacts on satellites create plasmas and surface damage, and a better understanding of the physics could provide insights applicable to damage mitigation \cite{meteorites1}.

\section{Can we accelerate progress?}                   

While good data and realistic models were published by 1905, and the problem is economically relevant, the modeling effort is not finished.  We believe that this effort requires a multi-disciplinary effort that considers accelerator, fusion, power transmission and many other applications of arcing to produce a coherent picture.  There are too many variables and mechanisms involved with arcing for a narrow approach to produce a useful, general result.  The basic experimental problem is that power supplies and cavities are usually inflexible, limiting the range of experiments.  A wilder variety of experiments, with better diagnostics would be welcome. 

Historically, support for these studies tended to insure that modeling was done with specific, narrow objectives.  This tended to produce a variety of models for a variety of initial conditions.  A better solution would be to encourage wider discussion among funding agencies and different communities to develop more general models.

The multi-disciplinary effort should be both experimental and theoretical.  It is interesting to note that Michelson and Millikan's experiments have not been repeated, in spite of their simplicity and utility.  With modern nanomanipulators, SEM systems, APT sample preparation and analysis, good electrical diagnostics and ion milling added to the 120 year old data, it would be possible to study many effects with both modeling and experimental data independent of application. Many of the uncertainties in arc behavior occur in the low power/energy stages of the arc, such as  breakdown and ionization, (Fig. 1), where experiments on fully developed arcs are less sensitive.  In addition, much of the modeling of the early stages of the arc is done at the atomic scale.  This work can be validated by experimental data at the nanometer level.  For example, it should be possible to measure the dependence of breakdown times and other effects on geometry, field polarity, DC/RF, and pulse length, to compare with predictions from modeling.   These studies should be generally relevant to many applications.

\end{document}